\newif\ifhyper
\hypertrue

\documentclass[conference]{IEEEtran}
\IEEEoverridecommandlockouts

\usepackage{cite}
\ifhyper
\RequirePackage{doi}
\else
\usepackage{url}
\fi

\usepackage{amsmath,amssymb,amsfonts}
\usepackage{algorithmic}
\usepackage{graphicx}
\usepackage{textcomp}
\usepackage{xcolor}
\usepackage{nicefrac}
\usepackage{multirow}
\usepackage{siunitx}
\usepackage{datetime}
\usepackage{numprint}
\usepackage{listings}

\npdecimalsign{.}
\npthousandsep{,}

\def\BibTeX{{\rm B\kern-.05em{\sc i\kern-.025em b}\kern-.08em
	T\kern-.1667em\lower.7ex\hbox{E}\kern-.125emX}}

\newcommand{\msf}[1]{\mathsf{#1}}

\newcolumntype{x}[1]{>{\centering\let\newline\\\arraybackslash\hspace{0pt}}p{#1}}
\newcolumntype{y}[1]{>{\raggedright\let\newline\\\arraybackslash\hspace{0pt}}p{#1}}
\newcolumntype{z}[1]{>{\raggedleft\let\newline\\\arraybackslash\hspace{0pt}}p{#1}}

\newcommand\Tstrut{\rule{0pt}{2.5ex}}		
\newcommand\Bstrut{\rule[-1ex]{0pt}{0pt}}	

\definecolor{xgray}{gray}{0.98}
\definecolor{xhl}{rgb}{1.0,1.0,0.8}
\definecolor{xgreen}{rgb}{0,0.3,0}
\definecolor{xblue}{rgb}{0,0,0.5}
\definecolor{xred}{rgb}{0.5,0,0}

\begin{document}

\title{A Lightweight ISA Extension for AES and SM4}

\author{\IEEEauthorblockN{Markku-Juhani O. Saarinen}%
\thanks{This work was supported by Innovate UK (R\&D Project Ref. 105747.)}\\
\textit{PQShield Ltd.}\\
	Oxford, United Kingdom \\
	{mjos@pqshield.com}
}

\maketitle

\makeatletter{\renewcommand*{\@makefnmark}{}
\footnotetext{\textit{\rule{8cm}{0.4pt} First International Workshop on Secure RISC-V Architecture Design Exploration  (SECRISC-V'20). It is held in conjunction with the IEEE International Symposium on Performance Analysis of Systems and Software (ISPASS) - August 23rd, 2020 in Boston, Massachusetts, USA.}}\makeatother}

\begin{abstract}
	We describe a lightweight RISC-V ISA extension for AES and SM4 block
	ciphers. Sixteen instructions (and a subkey load) is required to
	implement an AES round with the extension, instead of 80 without.
	An SM4 step (quarter-round) has 6.5 arithmetic instructions,
	a similar reduction. Perhaps even more importantly the ISA extension
	helps to eliminate slow, secret-dependent table lookups and to protect
	against cache timing side-channel attacks. Having only one S-box,
	the extension has a minimal hardware size and is well suited for
	ultra-low power applications. AES and SM4 implementations using the ISA
	extension also have a much-reduced software footprint. The AES and SM4
	instances can share the same data paths but are independent in the
	sense that a chip designer can implement SM4 without AES and vice versa.
	Full AES and SM4 assembler listings, HDL source code for instruction's
	combinatorial logic, and C code for emulation is provided to the
	community under a permissive open source license.
	The implementation contains depth- and size-optimized joint AES and
	SM4 S-Box logic based on the Boyar-Peralta
	construction with a shared non-linear middle layer, demonstrating
	additional avenues for logic optimization. The instruction logic has
	been experimentally integrated into the single-cycle execution path of
	the ``Pluto'' RV32 core and has been tested on an FPGA system.
\end{abstract}


\section{Introduction}

	The Advanced Encryption Standard (AES) is a 128-bit block cipher with
	128/192/256 - bit key, defined in the FIPS 197 standard \cite{NI01}.
	AES is a mandatory building block of the TLS 1.3 \cite{Re18} security
	protocol and is widely used for storage encryption, shared-secret
	authentication, cryptographic random number generation, and in many
	other applications.

	The SM4 block cipher \cite{SA16A} fulfills a similar role to AES in
	the Chinese market and is the main block cipher recommended for use
	in China. It is also standardized with ISO \cite{IS18}.
	SM4 also has a 128-bit block size, but only one key size, 128
	bits. Even though its high-level structure differs completely from
	AES, the two share significant similarities in their sole nonlinear
	component, which is a single $8 \times 8$-bit ``S-Box'' substitution
	table in both cases.

	Cache timing attacks on AES became well known in the mid-2000s when
	it was demonstrated that common table-based implementations can be
	exploited even remotely \cite{OsShTr06,Be05}; very similar issues
	also affect SM4. In presence of a cache, the only way to make the
	execution time of these ciphers fully independent of secret data is to
	eliminate the table lookup either by implementing it as bitsliced Boolean
	logic or by providing a specific ISA extension for the S-Box lookup.

	Consumer CPUs have had instructions to support AES for almost a
	decade via the Intel AES-NI in x86 \cite{Gu09} and ARMv8-A cryptographic
	extensions \cite{AR19}; these are almost universally available in PCs
	and higher-end mobile devices such as phones. ARM also supports SM4 via
	the ARMv8.2-SM extension. The AES instructions have been shown to make
	AES much less of a throughput bottleneck for high-speed TLS communication
	(servers) and storage encryption (mobile devices), thereby also extending
	battery life in the latter. Both Intel and ARM cryptographic
	ISAs require 128-bit (SIMD) registers and are not available on
	lower-end CPUs.

	In this work, we show that it is possible to create a simple AES and SM4
	ISA extension that offers a significant performance improvement
	and timing side-channel resistance with a minimally increased hardware
	footprint. It is especially suitable for lightweight RV32 targets.

\section{A Lightweight AES and SM4 ISA Extension}

	The ISA extension operates on the main register file only, using two
	source registers, one destination register, and a 5-bit field
	\verb|fn[4:0]| which can be seen either as an ``immediate constant''
	or just code points in instruction encoding.
	In either case, the interface to the (reference) combinatorial logic is:
\begin{lstlisting}
module saes32(
  output [31:0] rd,	  // to output register
  input	 [31:0] rs1,  // input register 1
  input	 [31:0] rs2,  // input register 2
  input	 [4:0]	fn	  // 5-bit func specifier
);
\end{lstlisting}
	See Section \ref{sec:interface} for encoding details of SAES32 as an
	RV32 R-type custom instruction for testing purposes. For RV64 the
	words are simply truncated or zero-extended.

	The five bits of $\verb|fn|$ cover encryption, decryption, and key
	schedule for both algorithms. Bits \verb|fn[1:0]| first select
	a single byte from \verb|rs2|. Two bits \verb|fn[4:3]| indicate which
	$8 \to 8$ - bit S-Box is used (AES, AES$^{-1}$, or SM4), and
	additionally \verb|fn[4:2]| specifies a $8 \to 32$ - bit linear
	expansion transformation (each of three S-Boxes has two alternative
	linear transforms, indicated by \verb|fn[2]|). The expanded 32-bit
	value is then rotated by 0--3 byte positions based on \verb|fn[1:0]|.
	The result is finally XORed with \verb|rs1| and written to \verb|rd|.

	Table \ref{tab:fnids} contains the identifiers (pseudo instructions)
	that we currently use for bits \verb|fn[4:2]|. We may arrange
	computation so that \verb|rd| = \verb|rs1| without increasing
	instruction count, making a two-operand ``compressed'' encoding possible.

	\begin{table}
	\caption{High-level assembler identifiers for fn[4:2].}
	\label{tab:fnids}
	\begin{center}
	\begin{tabular}{l c l}
	{\bf Instruction} & {\bf fn[4:2]} & {\bf Description or Use}	\\
	\hline
	\verb|saes32.encsm| & \verb|3'b000| & AES Encrypt round.		\\
	\verb|saes32.encs|	& \verb|3'b001| & AES Final / Key sched.	\\
	\verb|saes32.decsm| & \verb|3'b010| & AES Decrypt round.		\\
	\verb|saes32.decs|	& \verb|3'b011| & AES Decrypt final.		\\
	\verb|ssm4.ed|		& \verb|3'b100| & SM4 Encrypt and Decrypt.	\\
	\verb|ssm4.ks|		& \verb|3'b101| & SM4 Key Schedule.			\\
	{\it Unused}		& \verb|3'b11x| & ($4 \times 6=24$ points used.)\\
		\hline
	\end{tabular}
	\end{center}
	\end{table}

	For AES the instruction selects a byte from \verb|rs2|, performs a single
	S-box lookup ({\it SubBytes} or its inverse), evaluates a part of the MDS
	matrix ({\it MixColumns}) if that linear expansion is step selected,
	rotates the result by a multiple of 8 bits
	({\it ShiftRows}), and XORs the result with \verb|rs1| ({\it AddRoundKey}).
	There is no need for separate instructions for individual steps of AES
	as small parts of each of them have been incorporated into a single
	instruction. We've found that each one of these substeps requires
	surprisingly little additional logic.

	For SM4 the instruction has the same data path with byte selection,
	S-Box lookup, and two different linear operations, depending on whether
	encryption/decryption or key scheduling task is being performed.

	Both AES \cite{NI01} and SM4 \cite{SA16A} specifications are written
	using big-endian notation while RISC-V uses primarily little-endian
	convention \cite{_WaAs19}. To avoid endianness conversion the
	linear expansion step outputs have a flipped byte order.
	This is less noticeable with AES, but the 32-bit word rotations of SM4
	become less intuitive to describe (while wiring is equivalent).

	We refer to the concise reference implementation discussed in Section
	\ref{sec:refimpl} for details about specific logic operations required to
	implement the ISA extension, and for standards-derived unit tests.

\section{Using the AES and SM4 Instructions}

	AES and SM4 were originally designed primarily for 32-bit software
	implementation. SAES32/SSM4 adopts the ``intended'' 32-bit
	implementation structure but removes table lookups and rolls several
	individual steps into the same instruction. Both AES and SM4
	implementations are also realizable with the reduced ``E''
	register file without major changes.

\subsection{AES Computation and Key Schedule}

	The structure of an AES implementation is similar to a ``T-Table''
	implementation, with sixteen invocations of \verb|saes32.encsm|
	per round and not much else (apart from fetching the round subkeys).
	In practice, two sets of four registers are used to store
	the state, with one set being used to rewrite the other, depending
	on whether an odd or even-numbered round is being processed. AES
	has $r \in \{10, 12, 14\}$ rounds, depending on the key size
	which can be $\{128, 192, 256\}$, respectively. The final round requires
	sixteen invocations of \verb|saes32.encs|. The same instructions are
	also used in the key schedule which expands the secret key to
	$4r+4$ subkey words.

	The inverse AES operation is structured similarly, with 16
	\verb|saes32.decsm| per main body round and 16 \verb|saes32.decs| for the
	final round. These instructions are also used for reversing the
	key schedule.
	Four precomputed subkey words must be fetched in each round, requiring
	four loads (lw instructions) in addition to their address increment
	(typically every other round).
	There is no need for separate {\it AddRoundKey} XORs as the
	subkeys simply initialize either one of the four-register sets used
	to store the state.

	It is also possible to compute the round keys ``on the fly'' without
	committing them to RAM. This may be helpful in certain types of security
	applications. The overhead is roughly 30\%. However, if the load
	operation is much slower than register-to-register arithmetic,
	the overhead of on-the-fly subkey computation can become negligible.
	 On-the-fly keying is more challenging in reverse.

\subsection{SM4 Computation and Key Schedule}

	SM4 has only one key size, 128 bits. The algorithm has 32 steps,
	each using a single 32-bit subkey word. The steps are typically organized
	into 8 full rounds of 4 steps each.
	Due to its Feistel-like structure, SM4 does not require an inverse S-Box
	for decryption like AES, which is a substitution-permutation network (SPN).
	The inverse SM4 cipher is equivalent to the forward cipher, but with a
	reversed subkey order.

	Each step uses all four state words and one subkey word as inputs,
	replacing a single state word. Since input mixing is built from XORs,
	some of the temporary XOR values are unchanged and can be shared between
	steps. Each round requires ten XORs in addition to sixteen
	\verb|ssm4.ed| invocations, bringing the total number of arithmetic
	instructions to 26 per round -- or 6.5 per step. Therefore SM4 performance
	is slightly lower than that of AES-128, despite having fewer full rounds.

	The key schedule similarly requires 16 invocations of \verb|ssm4.ks|
	and 10 XORs to produce a block of four subkey words. The key schedule
	uses 32 ``CK'' round constants which can be either fetched from a table
	or computed with 8-bit addition operations on the fly.

	For SM4 each block of four consecutive invocations
	of \verb|ssm4.ed| and \verb|ssm4.ks| share the same source and
	destination registers, differing only in \verb|fn[1:0]| which steps
	through $\{0,1,2,3\}$. We denote such a four-SSM4 blocks as pseudo
	instruction \verb|ssm4.ed4| and \verb|ssm4.ks4|. One can reduce the
	per-round instruction count of SM4 from 26 (+4 lw) to 14 (+4 lw) 
	by implementing \verb|ssm4.ed4| as a ``real''  instruction. In terms
	of hardware area \verb|ssm4.ed4| would be almost four times larger than
	\verb|ssm4.ed| as it has four parallel S-Boxes.

	Note that a 32-bit ``T-Table'' type AES implementation does
	{\it not} benefit from four parallel S-Boxes in encryption or decryption,
	only in key schedule.

\section{Reference Implementation}
\label{sec:refimpl}

	An open-source reference implementation is available\footnote{AES/SM4
	ISA Extension: \url{https://github.com/mjosaarinen/lwaes_isa}}.
	The distribution contains HDL combinatorial logic for the SAES32
	instruction (including the S-Boxes) and provisional assembler listings
	for full AES-128/192/256 and SM4-128.

	The package also has (C-language) emulator code for the instruction
	logic, ``runnable pseudocode'' implementations of algorithms, and a
	set of standards-derived unit tests.
	This research distribution is primarily intended for obtaining data such
	as instruction counts and intermediate values but can be readily
	integrated into many RISC-V cores and emulators.

\subsection{About the AES, SM4 S-Boxes}

	AES and SM4 can share data paths so it makes sense to explore their
	additional structural similarities and differences.
	Both SM4 and AES S-Boxes are constructed from finite field inversion
	$x^{-1}$ in $\msf{GF}(2^8)$ together with a linear (affine)
	transformations on input and/or output. The inversion makes them
	``Nyberg S-Boxes'' \cite{Ny93} with desirable properties against
	differential and linear cryptanalysis, while the linear mixing steps
	are intended to break the bytewise algebraic structure.

	Since $x^{-1}$ is an involution (self-inverse) and affine isomorphic
	regardless of polynomial basis, AES, AES$^{-1}$, and SM4 S-Boxes really
	differ only in their inner and outer linear layers.

	Boyar and Peralta \cite{BoPe12} show how to build low-depth circuits
	for AES that are composed of a linear top and bottom layers and
	a shared nonlinear middle stage.
	Here XOR and XNOR gates are ``linear'' and the shared nonlinear layer
	consists of XOR and AND gates only. We created new outer layers for SM4
	that use the same the middle layer as AES and AES$^{-1}$.

	\begin{table}
	\caption{Algebraic gate counts for a Boyar-Peralta type low-depth
		S-Boxes that implement SM4 in addition to AES and AES$^{-1}$.}
	\label{tab:sboxgates}
	\begin{center}
	\begin{tabular}{l r@{\hskip3pt}c@{\hskip3pt}l z{8mm} z{8mm} z{8mm} z{8mm}}
	{\bf Component} & \multicolumn{3}{c}{In, Out}
		& \scriptsize{\sf XOR}	& \scriptsize{\sf XNOR}
		& \scriptsize{\sf AND}	& {\bf Total}	\\
	\hline
	Shared middle		& 21 & $\to$ & 18 & 30	& -		& 34	& 64	\\
	AES top				& 8	 & $\to$ & 21 & 26	& -		& -		& 26	\\
	AES bottom			& 18 & $\to$ & 8  & 34	& 4		& -		& 38	\\
	AES$^{-1}$ top		& 8	 & $\to$ & 21 & 16	& 10	& -		& 26	\\
	AES$^{-1}$ bottom	& 18 & $\to$ & 8  & 37	& -		& -		& 37	\\
	SM4 top				& 8	 & $\to$ & 21 & 18	& 9		& -		& 27	\\
	SM4 bottom			& 18 & $\to$ & 8  & 33	& 5		& -		& 38	\\
	\hline
	\end{tabular}
	\end{center}
	\end{table}

	Each S-Box expands an 8-bit input to 21 bits in a linear inner (``top'')
	layer, uses the shared nonlinear 21-to-18 bit mapping as a middle
	layer, and again compresses 18 bits to 8 bits in the outer (``bottom'')
	layer. Table \ref{tab:sboxgates} gives the individual gate counts
	to each layer; summing up top, middle, and bottom gives the total
	S-Box gate count ($\approx$ 128).

	Despite such a strict structure and limited choice of gates
	(that is suboptimal for silicon but very natural to
	mathematics), these are some of the smallest circuits for AES known.
	Note that it is possible to implement AES with fewer gates (113 total),
	but this results in 50\% higher circuit depth \cite{BoPe10}.

\subsection{Experimental Instruction Encoding and Synthesis}
\label{sec:interface}

	For prototyping we interfaced the SAES32 logic using the {\it custom-0}
	opcode and R-type instruction encoding with \verb|fn[4:0]| occupying
	lower 5 bits of the funct7 field:

	\begin{scriptsize}
	\begin{center}
	\begin{tabular}{| c | c | c | c | c | c | c | }
		\multicolumn{1}{c}{[31:30]} &
		\multicolumn{1}{c}{[29:25]} &
		\multicolumn{1}{c}{[24:20]} &
		\multicolumn{1}{c}{[19:15]} &
		\multicolumn{1}{c}{[14:12]} &
		\multicolumn{1}{c}{[11:7]}	&
		\multicolumn{1}{c}{[6:0]}	\\
	\cline{1-7}
	00 & {\tt fn} & {\tt rs2} & {\tt rs1} & 000 & {\tt rd} & 0001011
		\Tstrut\Bstrut\\
	\cline{1-7}
	\end{tabular}
	\end{center}
	\end{scriptsize}

	The implementation has been tested with PQShield's ``Pluto'' RISC-V
	core. We synthesized the same core on low-end Xilinx Artix-7 FPGA target
	(XC7A35TICSG324-1L) with and without the SAES32 (AES, SM4) instruction
	extension and related execution pipeline interface. Table \ref{tab:socsize}
	summarizes the relative area added by SAES32. For comparison, we also
	measured the size of a memory-mapped AES module ``EXTAES''. This module
	implements AES encryption only.

	Based on our FPGA experiments we estimate that the full (AES, AES$^{-1}$,
	SM4) instruction proposal increases the amount of core logic (LUTs)
	by about 5\% over a typical baseline RV32I core, but relatively much
	less for more complex cores.

	Table \ref{tab:yosys} contains area estimates for the SAES32 module
	(not the additional decoding logic)
	using the Yosys ``Simple CMOS'' flow which uses a mock-up ASIC
	cell library. Here GE is the gate count (NAND2 Equivalents) and Longest
	Topological Path (LTP) is the reported depth (delay) measure.

	Implementors can experiment if it is beneficial to multiplex the S-Box
	linear layers with the shared middle layer. The required mux logic seems
	large and increases the circuit depth, so our current reference
	implementation does not use it.


	\begin{table}
	\begin{center}
	\caption{RV32 SoC area with and without SAES32 (AES, AES$^{-1}$, SM4);
		``Pluto'' core on an Artix-7 FPGA. EXTAES is a CPU-external
		memory-mapped AES-only module, presented for comparison.}
	\label{tab:socsize}
	\begin{tabular}{| l | r	 | r@{\hskip3pt}l  | r@{\hskip3pt}l |}
		\hline
		{\bf Resource}	& {\bf Base}
		& \multicolumn{2}{c|}{{\bf SAES32}	($\Delta$)}
		& \multicolumn{2}{c|}{{\bf EXTAES} ($\Delta$)} \\
		\hline
		Logic LUTs	& 7,767 & 8,202 & (+435) & 9,795 & (+2,028	\\
		Slice regs	& 3,319 & 3,342 & (+23)	 & 4,361 & (+1,042) \\
		SLICEL		& 1,571 & 1,864 & (+293) & 2,068 & (+497)	\\
		SLICEM		&	734 &	737 & (+3)	 &	 851 & (+117)	\\
		\hline
	\end{tabular}
	\end{center}
	\end{table}

	\begin{table}
	\begin{center}
	\caption{Yosys Simple CMOS Flow area estimates for SAES32.}
	\label{tab:yosys}
	\begin{tabular}{| l | r	 | r  | r |}
	\hline
	{\bf Target}		& GE (NAND2)	& Transistors	& LTP	\\
	\hline
	AES Encrypt only	& 642			& 2,568			& 25	\\
	SM4 Full			& 767			& 3,066			& 25	\\
	AES Full			& 1,240			& 4,960			& 28	\\
	AES + SM4 Full		& 1,679			& 6,714			& 28	\\
	\hline
	\end{tabular}
	\end{center}
	\end{table}

\section{Performance and Security Analysis}

	The hand-optimized AES implementation\footnote{Ko Stoffelen:
	``RISC-V Crypto'' \cite{St19} \url{https://github.com/Ko-/riscvcrypto}}
	referenced in \cite{St19} requires 80 core arithmetic instructions per
	round. The same task can be accomplished with 16 SAES32 instructions.
	Furthermore, 16 of those 80 are memory loads, which typically require
	more cycles than a simple arithmetic instruction (or SAES32). Each AES
	round additionally requires a few operations for loading subkeys and
	managing instruction flow.

	Overall, based on RV32 and RV64 instruction counts we estimate that the
	performance of a SAES32 AES can be expected to be more than 500\% better
	than the fastest AES implementations that use the baseline ISA only.
	Much of the precise performance gain over a table-based
	implementation depends on the latency of memory load operations.

	SAES32-based AES and SM4 implementations are inherently constant-time and
	resistant to timing attacks. Stoffelen \cite{St19} also presents a
	constant-time, bitsliced AES implementation for RISC-V which requires
	$2.5$ times more cycles than the optimized table-based implementation.
	So SAES32 speedup over a timing side-channel hardened base
	ISA implementation is expected to be roughly 15-fold.

	We are not aware of any definitive assembler benchmarks for SM4 on
	RISC-V, but based on instruction count estimates the performance
	improvement can be expected to be roughly similar or more (over 500 \%).
	Without SAES32 a simple SM4 software implementation would benefit
	from rotation instructions (in the proposed RISC-V bit
	manipulation extension).

	We have only discussed timing side-channel attacks. Since these
	instructions interact with the main register file, any electromagnetic
	emission countermeasures would probably have to be extended to additional
	parts of the CPU core.

\section{Conclusions}

	We propose a minimalistic RISC-V ISA extension for AES and SM4
	block ciphers. The resulting speedup is 500\% or more for both ciphers
	when compared to hand-optimized base ISA assembler implementations
	that use lookup tables.

	In addition to saving energy and reducing latency in secure communications
	and storage encryption, the main security benefit of the instructions
	is their constant-time operation and resulting resistance against cache
	timing attacks. Such countermeasures are expensive in pure software
	implementations.

	The instructions require logic only for a single S-Box, which is combined
	with additional linear layers for increased code density and performance.
	The hardware footprint of the instruction is very small as a result.
	If both AES and SM4 are implemented on the same target they can share
	data paths which further simplifies hardware. However, AES and SM4
	are independent of each other and AES$^{-1}$ is also optional.
	It is not rare to implement and use the forward AES without inverse AES
	as common CTR-based AES modes (such as GCM) do not require the inverse
	cipher for decryption \cite{Dw01}.

	This proposal is targeted towards (ultra) lightweight MCUs
	and SoCs. A different type of ISA extension may provide additional
	speedups on 64-bit and vectorized platforms, but with the cost of
	increased implementation area. Designers may still want
	to choose this minimal-footprint option if timing side-channel
	resistance is their primary concern.

	{\bf Postscript.} Since the original preprint distribution
	of this work in February 2020, these 32-bit scalar	AES and SM4
	instructions have been evaluated (AES as ``$\nu_3$'' in \cite{MaNePa+20})
	and adopted into the RISC-V Crypto Extension Proposal \cite{_Ma20}.
	We have changed the instruction naming (from ``ENC1S'' to ``SAES32'')
	in this paper to reflect that specification.

\ifhyper
\bibliographystyle{abbrvurl}
\else
\bibliographystyle{abbrv}
\fi
\bibliography{lwaes}

\end{document}